# Junction-less nanowire based photodetector: Role of nanowire width

Anita Fadavi Roudsari[1†], Iman Khodadadzadeh[1], Nixon O[2], Simarjeet S. Saini[1], and M. P. Anantram[3,*]

[1]*Department of Electrical and Computer Engineering, University of Waterloo, Waterloo, Ontario N2L 3G1, Canada*
[2]*Teledyne DALSA Inc., 605 McMurray Rd, Waterloo, Ontario, Canada*
[3]*Department of Electrical Engineering, University of Washington, Seattle, Washington 98105-2500, USA*
*\*afadavi@uwaterloo.ca, anant@uw.edu*

[†]*Current Address: Department of Microtechnology and Nanosciense - MC2, Chalmers University of Technology, SE-412 96 Göteborg, Sweden*

**Abstract**

Enhanced photocurrent is demonstrated in a junction-less photodetector with nanowires embedded in its channel. The fabricated photodetector consists of a large area for efficient absorption of incident light with energy band engineering achieved in nanowires. The structure design consists of a set of two symmetrically positioned gates, primary and secondary, that are located over the nanowires. Each gate is used for biasing and control of the charge flow. We find that detectors with narrower nanowires controlled by their secondary gate generate larger photocurrents under similar illumination conditions. Our results show that while the dark current remains the same, the photocurrent increases as the nanowire width decreases.

## 1. Introduction

One can find several reasons behind incorporating nanowires in the design of photodetector devices. Some semiconductors such as silicon and germanium show change in bandgap as their size is reduced towards quantum mechanical limits. This can in turn change their optical properties [1], resulting in better absorption in nanowire based photo-devices.

The second reason is the advantage of packing nanowires in arrays [2-5]. This increases the fill factor of the photodetector, especially in vertically arranged wires, resulting in efficient light trapping due to resonant excitement of transverse optical modes [6]. The array configuration helps to compensate the poor ability of a single nanowire in absorbing light [7].

Thirdly, nanowires can be incorporated in channels of metal-oxide-semiconductor photodetectors. Due to their small cross sectional area the electric field of the gate controls the charge transport within the nanowire channel in a more effective way compared with bulk semiconductor devices [8, 9]. This configuration enables a metal oxide semiconductor (MOS) transistor to operate under lateral bipolar action [10]. In a subcategory of such detectors the device itself is junction-less, and it is the energy band control by the gate which creates an artificial junction [11-12]. Such devices work under lateral bipolar action without the need for having physically doped junctions [13-17].

Recently we introduced the design concepts of a junction-less photodetector in which the nanowires are not used as the main absorption medium [18]. Instead, the design makes use of a wide area in the channel to absorb light efficiently. Two nano-scale regions are also part of the channel, each controlled by a gate. While the first gate/ nanowire combination is used to create the junction, the second combination is incorporated to increase the photocurrent. In this work, we present the results of the experimentally fabricated photodetectors that verify the importance of using nano-scale channels discussed in [18].

## 2. Experimental: Device Fabrication and Measurements

Devices were fabricated on p-type silicon on Insulator (SOI) wafers, with top silicon layer of 35nm. The resistivity of the silicon layer was between 8.5 and 22 Ωcm. The thickness of the buried oxide layer was 145±5nm. Mesas of silicon for each device were fabricated by electron beam (e-beam) lithography and dry etching, where a negative resist (Ma-N 2400 series) was used for patterning and also as the etching mask. The samples were then cleaned by Piranha process, and dipped in Buffered hydro fluoric acid (BHF) mixture for native oxide removal. Next, each sample was covered with a layer of 25nm thick Aluminum Oxide, deposited by plasma assisted atomic layer deposition at 300°C. Source, drain and gate contacts were fabricated by e-beam lithography, dielectric removal (source and drain only), deposition of metal (Aluminum, 250nm), and the lift off process. At the end, the samples were annealed in forming gas at 425°C for 30 minutes.

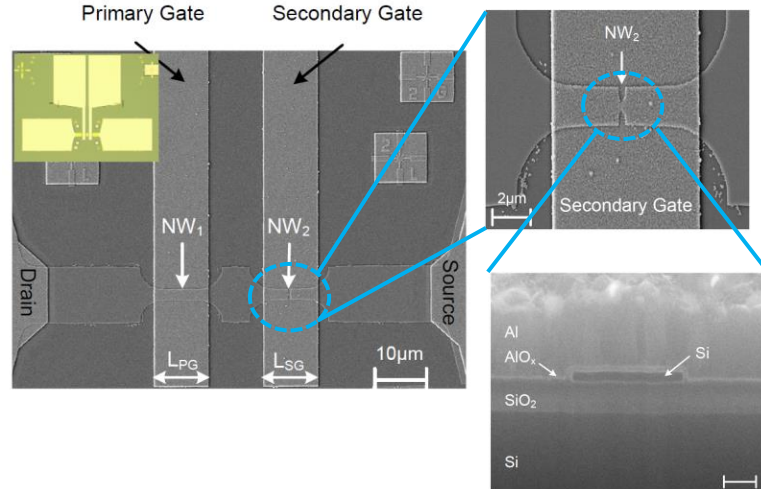

Figure 1 shows the Scanning Electron Micrograph (SEM) image of a representative fabricated structure. Each photodetector is a four contact device, consisting of source, drain and two gates that are located symmetrically with respect to the source and drain. The middle, wide absorption region is surrounded by two nanowires, $NW_1$ and $NW_2$. We studied different widths of $NW_1$ and $NW_2$ as discussed later. For the device imaged, $NW_1$ is approximately 2µm wide and 10µm long, while $NW_2$ is 300nm wide at its narrowest. The primary and secondary gates cover the $NW_1$ and $NW_2$ regions, respectively. They are both 10µm long for all of the results presented in this manuscript. The inset in figure 1 shows the cross section of the device, across $NW_2$.

Figure 1: Junction-less photodetector with multiple gates. The primary gate (PG) covers $NW_1$, while the secondary gate (SG) covers $NW_2$. $L_{PG}=L_{SG}=10\mu m$. The top inset shows a larger view of $NW_2$. The bottom inset is the cross section of $NW_2$. The scale bar is 200nm.

The operation principles of junction-less photodetectors with multiple gates are reviewed briefly [18]. Compared with the junction-less transistors of reference [11], one major difference in our design is reshaping the channel for light absorption purpose [18]. As shown in figure 1, the area between the source and drain contacts is designed to be partly wide to efficiently absorb light. The gate/nanowire combination in the two regions of $NW_1$ and $NW_2$ are designed to control the operation of the device. The primary gate is responsible for reducing the dark current and creating the lateral bipolar action [10]. Similar to [11], when biased positively, the primary gate depletes the $NW_1$ area out of majority carriers (holes, p-type semiconductor) and creates a potential barrier that prevents the charges from easily drifting towards the contacts. As plotted in figure 2(a) this behavior is similar to a bipolar transistor with both junctions reversed biased. The high potential barrier causes the dark current to be low, as illustrated in Figure 2(b). During illumination, the electric field created by the bias of the contacts forces the photo-generated majority carriers (now, electrons in the channel) to drift towards the contacts; while the minority carriers get trapped inside the barrier. As shown in figure 2(a), the trapped carriers lower the potential barrier; and therefore, the overall charge flow during illumination is increased (figure 2(b)).

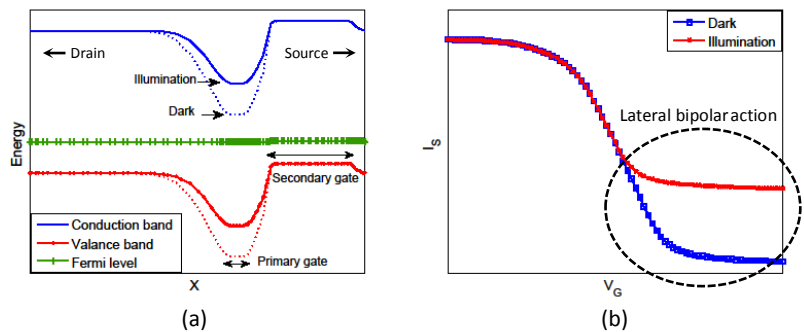

Figure 2: (a) Energy band diagram of a junction-less photodetector along drain-channel-source, under dark (dashed) and illumination by light (solid). $V_{G(primary)} > 0$, $V_{G(secondary)} < 0$, and $V_S=V_D=0$. (b) Source current as a function of the primary gate bias, showing the lateral bipolar action when $V_{G(primary)} > 0$. $V_D =0$, and $V_s > 0$.

The second major difference of our device with the one in [11] is introducing a secondary gate to the channel to increase the photocurrent. The secondary gate controls the potential barrier of the $NW_2$ region. This gate is designed to bias $NW_2$ in accumulation; to pull up the potential barrier in this region. For a p-type device, the negatively biased secondary gate increases the carrier concentration (holes) on the source area; and causes the emitter efficiency of the device to increase, resulting in an increased photocurrent. The importance of the secondary gate in controlling the carrier concentration is discussed in [18]. The potential barrier on the source region drops in the absence of the secondary gate.

The width of $NW_1$ and $NW_2$ plays an essential role in enabling the gates to effectively control the channel. Among the two channel regions, for the devices presented here the width of $NW_1$ is less critical. This is due to the fact that the 35nm layer of silicon in the fabricated structures is thin enough to be fully depleted by the primary gate. However, biasing in accumulation by the secondary gate is a more localized effect that mostly influences the charges that are close to the interface of the semiconductor and dielectric. As a result, as $NW_2$ gets narrower, a larger volume of $NW_2$ is affected by the secondary gate, and one expects a larger photocurrent in the device with a narrower $NW_2$.

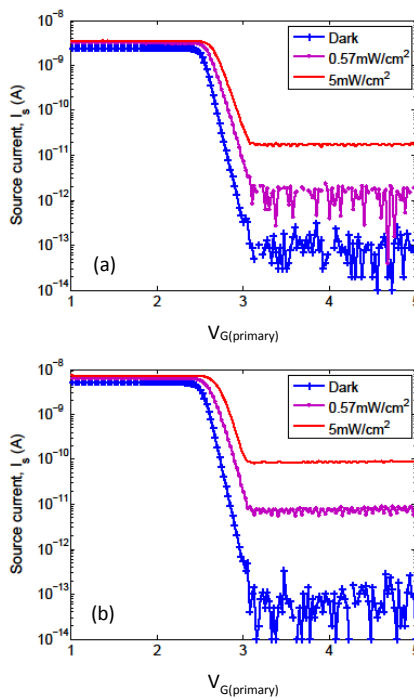

In order to illustrate the role of $NW_2$/ secondary gate combination, the currents of two fabricated devices are compared in figures 3(a) and (b) for devices with $NW_2$ widths of 2 μm and 300 nm respectively. Each figure represents the current as a function of the primary gate bias, under dark and also for two illumination intensities, where the top side of the device is illuminated. The width of $NW_2$ differs in the two devices, but other dimensions of the detectors are similar. At $V_{G(primary)}$ of greater than about 3.25V, where the photodetectors operate under lateral bipolar action, both devices show an equal level of dark current below 0.1 pA. However, under illumination and for both intensities, despite being about 6.5 times narrower, the photocurrent of the structure with $NW_2$:300nm wide (bottom) is more than 4.5 times larger than that of the device with $NW_2$: 2μm wide (top). This is an actual enhancement achieved in the terminal current; not current per unit area as is normally reported for nanowire devices.

During photocurrent measurements, in addition to the alignment considerations, we perform another experiment to ensure that the observed results are not due to measurement setup errors such as misplacement or poor coupling of the light source, or due to variances in the fabricated devices. This time, during the whole experiment, the light source and also the device are kept at fixed positions. First, the dark and illumination currents are measured while $V_D$ =0V, $V_S$ =1V, and $V_{G(secondary)}$ =-3V; and the primary gate voltage is varied between 1 and 5V. We name this experiment as the 'normal' measurement.

Figure 3: Current versus the primary gate bias for two photodetector devices with (a) $NW_1$:10μm, $NW_2$:2μm, and (b) $NW_1$:10μm, $NW_2$:300nm wide. $V_D$ =0V, $V_S$ =1V, and $V_{G(secondary)}$ =-3V. The device is illuminated on the top by a laser diode with the wavelength of λ=405nm. An Agilent 4155C parameter analyzer is used for current/ voltage measurements.

We compare the above 'normal' measurement to the 'mirrored' measurement. 'Mirrored' measurements refer to the condition where the roles of source and drain are flipped compared to 'normal' measurements, and the roles of the primary and secondary gates are also flipped. That is, now, $V_D$=1V, $V_S$=0V, $V_{G(primary)}$=-3V and $V_{G(secondary)}$ is varied between 1 and 5V. The nomenclature of 'primary' and 'secondary' gates is the same as figure 1.

This experiment is important as it can show the impact of the nanowire size more clearly using the same device. The effect of unwanted, external factors such as variation in the silicon thickness and the size of (nominally equal) channels in different devices, and also alignment errors during placement of the light source are minimized here. Furthermore, the symmetric position of the gates with respect to the source and drain regions allows for the specific comparison of the nanowires.

The normal/ mirrored measurement results for two devices are presented in figure 4, where the photocurrents are measured for two wavelengths of 405nm and 632nm. In agreement with our previous observations shown in figure 3, in each device, the illumination current is larger when the channel underneath the *negatively biased* gate is narrower.

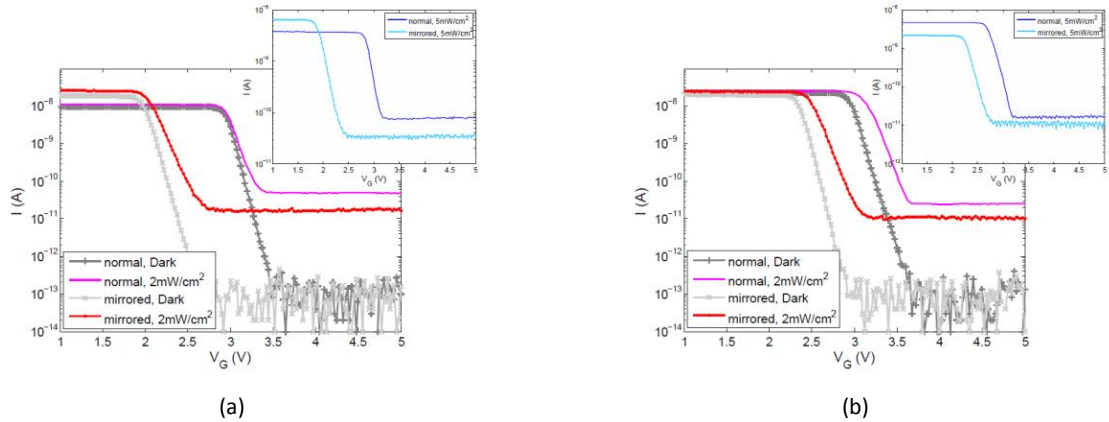

Figure 4: Normal measurement: change of the current (at source contact) as a function of primary gate bias, under dark and illumination. $V_D$=0V, $V_S$=1V, $V_{G(secondary)}$=-3V and λ=632nm. Mirrored measurement: change of the current (at drain contact) as a function of secondary gate bias (acting as primary gate), under dark and illumination. $V_D$=1V, $V_S$=0V, and $V_{G(primary)}$=-3V. (a) $NW_1$:10μm and $NW_2$:300nm wide. (b) $NW_1$:10μm, and $NW_2$:2μm wide. Insets represent the normal and mirrored photocurrents when λ=405nm.

We present modeling results to further discuss the device response under normal and mirrored measurements using two dimensional simulations, and drift-diffusion framework [19]. The conduction band energy of a multiple gate structure under normal and mirrored biasing is plotted in figures 5(b) and (c). The device itself, shown in figure 5(a), has the same channel and primary/ secondary gate length as the fabricated devices. However, $NW_1$ in this structure is 500nm wide; while $NW_2$ is 40nm. The main reason for this change comes from the fact that in the fabricated structure, the 2 μm wide, 35 nm thick $NW_1$ is depleted by the top side of the primary gate. However, in a two dimensional device with $NW_1$ of 2 μm width, the primary gate is unable to deplete the channel, and the results cannot convey our primary goal. The change in the nanowire widths will cause some difference in the resulting currents, but does not limit our goal of qualitatively and effectively discussing how the nanowires operate in the fabricated device.

The energy bands are obtained along the channel, shown by AA' in figure 5 (a). The energy bands show the importance of the nanowire width under the negatively biased gate. Under 'normal' biasing, the secondary gate strongly controls the barrier height of $NW_2$ (40μm <x<50μm) as shown in figure 5(b). In comparison, under 'mirrored' biasing, the primary gate is not as effective in controlling the barrier height on $NW_1$ (20μm<x<30μm) as shown in figure 5(c). Therefore, under illumination, the change in barrier height for 'normal' biasing (20µm<x<30µm of figure 5(b)) is larger than the change in barrier height for 'mirrored' biasing (40µm<x<50µm of figure 5(c)). This causes the photocurrent under normal biasing to be larger, as illustrated in figure 5(d).

We also notice from figure 5(b) and (c) that under dark, the *overall* barrier height (for the charge flow) is larger when the biasing is mirrored. The reason is the stronger effect of the positively biased secondary gate in depleting the 40nm wide $NW_2$ (figure 5(c), 40µm<x<50µm), compared to the case of 500nm channel when the device is biased normally (figure 5(b), 20µm<x<30µm). As a result, as plotted in figure 5(d), we expect the level of dark current under mirrored biasing to be smaller than that of the normal case. The reason that the dark currents in figure 4 are at the same level is due to accuracy limitation of our measurement setup.

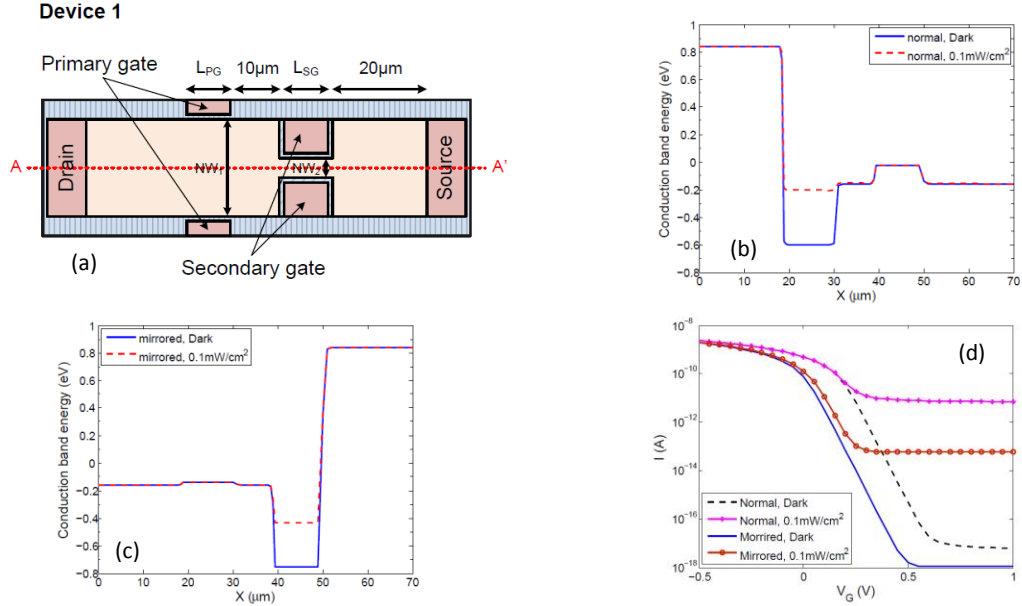

Figure 5: Results from modeling; (a) Two dimensional, multiple gate junction-less structure with $NW_1$:0.5µm, and $NW_2$:40nm wide. (b) Conduction band energy (cutline AA'), normal biasing ($V_D$ =0V, $V_S$ =1V, and $V_{G(secondary)}$ =-2V). (c) Conduction band energy (cutline AA'), mirrored biasing ($V_D$ =1V, $V_S$ =0V, and $V_{G(primary)}$ =-2V). (d) Current as a function of gate bias. λ=405nm. The current is scaled to demonstrate the 35 nm device thickness. Surface reflection is considered 50%.

We should emphasise that the fabricated devices are not optimized to generate a large gain. There are a number of parameters that can be modified in order to improve the device in figure 1 [18]. For example, in the fabricated structures the two gates are placed symmetrically, as it was necessary for us to be able to try normal and mirrored measurements. However, the symmetrical placement of the gates causes less photo-generated carriers to contribute to the lateral bipolar action. Furthermore, the 250nm thick aluminum that is used as the gate material reflects the incident light. To provide a wider area for light absorption, the length of the primary gate in figure 1 can be reduced.

Table 1 illustrates how such modifications can improve the photocurrent. In comparison with *Device 1* shown in figure 5(a), some changes are made in *Device 2*. The distance between the two gates is reduced to 5µm. The gates are shifted towards the source such that the gap between the secondary gate and the source is also 5µm. In addition, the width of the primary gate is reduced to 1µm. As a result, in comparison with Device 1, the photocurrent increases by more than 45 times. The dark current, and also the optical gain of the two devices are summarized in table 1 as well. The data shows the importance of design modifications in improving the optical gain.

Besides design considerations, we would also like to remark that surface recombination is another important parameter that can degrade a transistor's performance. In a phototransistor, a large surface recombination velocity increases the level of dark current, and decreases the photocurrent. The exposed silicon surface in our device is covered with a layer of aluminum oxide, and annealed at 425°C. Although this passivates the silicon dangling bonds and reduces the surface recombination [20-22], we believe better surface passivation can enhance the device performance. To clarify this, we modeled Device 1 and Device 2, assuming the surface recombination velocity (SRV) ranges from 10 to $10^4$ cm/s. The results are listed in Table 1. In comparison with the ideal case, both devices show degradation of performance as the surface recombination velocity increases, with Device 2 showing a faster trend, due to its larger optical gain. The 'ideal' optical gain of Device 2 is about 40 times larger than that of Device 1, meaning that the variation of barrier height at $NW_1$ in Device 2 is more sensitive to the number of trapped electrons. Therefore, recombination of the photo-generated electrons due to surface traps leads to a larger degradation in the current of Device 2. While our manuscript does not intend to address engineering interfaces, which we agree is a very important area being pursued by many researchers, the results in table 1 indicate that surface passivation must be addressed properly to further reveal the beauty and power of such nanowire based phototransistors.

Table 1: Role of device modification and effect of surface recombination on phototransistor response

|  | Device 1 | | | Device 2 | | |
| --- | --- | --- | --- | --- | --- | --- |
|  | $I_{dark}$ (A/μm) | $I_{photo}$(A/μm) | $G_{opt}$ | $I_{dark}$ (A/μm) | $I_{photo}$(A/μm) | $G_{opt}$ |
| **Ideal** | $1.8\times10^{-16}$ | $4.1\times10^{-10}$ | 50.9 | $1.3\times10^{-15}$ | $1.9\times10^{-8}$ | 2068.9 |
| **SRV=10cm/s** | $5.3\times10^{-16}$ | $9.96\times10^{-11}$ | 12.5 | $1.6\times10^{-15}$ | $3.8\times10^{-9}$ | 403.5 |
| **SRV=$10^2$cm/s** | $2.2\times10^{-15}$ | $1.5\times10^{-11}$ | 1.84 | $3.1\times10^{-15}$ | $3.2\times10^{-10}$ | 34.4 |
| **SRV=$10^3$cm/s** | $1.8\times10^{-14}$ | $3.4\times10^{-12}$ | 0.42 | $1.8\times10^{-14}$ | $1.4\times10^{-11}$ | 1.5 |
| **SRV=$10^4$cm/s** | $1.7\times10^{-13}$ | $1.5\times10^{-12}$ | 0.14 | $1.7\times10^{-13}$ | $1.6\times10^{-12}$ | 0.13 |

$NW_1$: 500nm, $NW_2$: 40nm.
$G_{opt}$: Optical gain, values are obtained for intensity of 0.1mW/cm$^2$.
$V_{G(primary)}$=1.0V, $V_{G(secondary)}$=-2.0V.
A surface reflection of 50% is considered. The photocurrent presented here is multiplied to transmission.
Minority carrier lifetime = $10^{-5}$s.

## 3. Conclusion

In summary, we demonstrated a novel design to incorporate nanowires in phototransistors that improves the performance of such detectors. In contrast to other designs in which nanowires serve as the absorption medium and also a part of the device, in our device the nanowires are mostly used for gain purpose. We fabricated junction-less photodetectors in which the charge flow is controlled by a set of two nanowire/gate combinations. The strong capacitive coupling of the gate and the nanowire allows for a strong control over the charge transport by the gate. Operation of the photodetector is based on the lateral bipolar action, by creating a potential barrier using the primary gate that leads to a reduced dark current. Incorporation of the secondary gate and a nanowire close to the source contact allows for increase of carrier concentration at the source, leading to a larger photocurrent. We highlighted the critical role of the secondary gate/ nanowire combination by demonstrating that the photocurrent can be enhanced if the width of the nanowire is reduced.

The results presented here verify our design considerations proposed in reference [18] that band engineering by using nanowire/ gate geometries can improve the photocurrent in phototransistors. We also discussed possible reasons for the low optical gain in the fabricated devices. Work to improve the surface passivation will increase the optical gain, leading to phototransistors with high optical gain.

**Acknowledgments**

The authors acknowledge support from National Science Foundation (NSF) grant #1001174, Ontario Centres of Excellence (OCE), Waterloo Institute for Nanotechnology (WIN) fellowship, and also access to the Institute for Quantum Computing (IQC, now Mark and Ophelia Lazaridis Quantum-Nano Centre) and the Centre for Advanced Photovoltaic Devices and Systems (CAPDS) cleanrooms.